\documentclass[preprint,12pt]{elsarticle}
\usepackage[utf8]{inputenc}
\usepackage{amsmath,amssymb,amsfonts}
\usepackage{algorithmic}
\usepackage{graphicx}
\usepackage{textcomp}
\usepackage{xcolor}
\usepackage{listings}
\usepackage{tabularx}
\usepackage{array}
\usepackage{xspace}
\usepackage{url}
\usepackage{textcomp}

\definecolor{codegreen}{rgb}{0,0.6,0}
\definecolor{codegray}{rgb}{0.5,0.5,0.5}
\definecolor{codepurple}{rgb}{0.58,0,0.82}
\definecolor{backcolour}{rgb}{0.95,0.95,0.92}
\begin{document}
\title{\textbf{Developing and Sustaining a Student-Driven Software Solutions Center - An Experience Report}}
\author[1]{Saheed Popoola \corref{cor1} \fnref{fn1}}
\ead{saheed.popoola@uc.edu}

\author[1]{Vineela Kunapareddi}

\author[1]{Hazem Said}

\cortext[cor1]{Corresponding author}
\fntext[fn1]{This is the first author footnote.}

\affiliation[1]{organization={School of Information Technology, University of Cincinnati},
            city={Cincinnati},
            state={Ohio},
            country={USA}}

\newcommand{\spa}[1]{{#1}}
\begin{abstract}
    Traditional approaches to software engineering education often limit students' exposure and engagement to real-world projects; thereby, failing to fully harness their potential and creativity. Yet, the dynamic and rapidly-advancing digital landscape means that there is a continuous need to empower students to become active participants, problem solvers, and innovators in delivering high-quality software solutions. Therefore, it is not surprising that fresh graduates are often ill-equipped to  handle industrial projects. Existing approaches to exposing students to industrial projects such as internships or capstone projects have not achieved the desired result because industries are often reluctant to assign important tasks to interns, and a capstone project is likely to be discontinued by the student after completing the course. Furthermore, all the team members in a capstone project are usually inexperienced engineers, and this may limit mentorship opportunities for the students.
    This paper presents an experience report on the establishment and sustenance of a student-driven software solutions center named \textit{Information Technology Solutions Center (ITSC)}, a unit within the School of Information Technology at the University of Cincinnati. A student-driven solution center empowers students to drive the design, development, execution, and maintenance of software solutions for industrial clients. This exposes the students to real-world projects and ensures that students are fully prepared to meet the demands of the ever-changing industrial landscape. The ITSC was established over a decade ago, has trained over 100 students, and executes about 20 projects annually with several industrial partners including Fortune 500 companies, government institutions, and  research agencies. This paper discusses the establishment and maintenance of the center with the goal of motivating and providing a clear blueprint for computing programs that want to establish a similar student-driven software solutions center.
\end{abstract}

\maketitle

\section{Introduction}

The software engineering field is rapidly evolving and the continuous need for high-quality software solutions that can target the unique needs of clients or users is evident. The need for custom software solutions also drives the demand for good software developers with sound software engineering methodologies. Hence, it is important for universities, which are seen as the beacon for training and learning, to efficiently train students that can build high-quality software. Unfortunately, current approaches to software engineering education often limit student engagement and creativity \cite{garousi2019aligning,garousi2019closing}. Furthermore, the software development industry has often noted that fresh graduates have limited exposure to real-world projects and are often ill-equipped to deal with industrial projects \cite{brechner2003, brng2013essence, radermacher2013, radermacher2014}.

 Two main approaches have been adopted to expose students to industrial projects with the goal of ensuring that students are adequately prepared to meet industry demands. The first approach is student internship where students are trained in an industry to work on software projects \cite{kapoor2019}. The internship often occurs when the school is on break, typically in the summer.  However, it is challenging to monitor student progress during the internship because the companies are often independent of the university. Furthermore, the short period of the internship (about 3 months) limits any meaningful contribution and students are assigned tasks with limited design problems \cite{shin2013}.
 
The second approach involves incorporating industrial projects in the student capstone projects \cite{keogh2007, tenhunen2023}. Unfortunately, many of the capstone projects often lack an industrial partner \cite{tenhunen2023}. Furthermore, the main goal of the student is to pass the course, while delivering high-quality software for the client is secondary. Hence, the project is likely to be discontinued at the completion of the course and there is no long-term support for the projects. Students are often taking other classes simultaneously, and it may be challenging for team members to meet regularly due to differing schedules. Furthermore, all the team members are almost inexperienced engineers and this may limit mentorship opportunities and peer-learning \cite{knudson2009}.

This paper explores the concept of a faculty-led and student-driven software solutions center where students drive the development and maintenance of customized software solutions for industrial clients such as profit and non-profit organizations, government institutions, research agencies, etc. The management of the center and sourcing for new projects might be handled by professionals or non-students. However, the implementation of the software projects would be led by the students.


A student-driven software solutions center can serve as a catalyst for innovation and collaborative problem-solving, enabling students to address real-world challenges through software development. These centers can provide students with the resources, mentorship, and opportunities to ideate, design, develop, and implement software applications or tools. By giving students the autonomy and support to explore their passions, apply their technical skills, and make a tangible impact, these centers foster a sense of ownership and agency among students, enhancing their learning experiences and preparing them for the demands of an increasingly digital world \cite{paasivaara2018does, marques2017enhancing}. Furthermore, the constant interaction of these centers with the industry can also facilitate the continuous updates and improvement to the software engineering course syllabus and curriculum in the universities.

The Information Technology Solutions Center (ITSC) \cite{itsc}, is a unit within the School of Information Technology at the University of Cincinnati that provides innovative, low-cost, and customized software solutions for industries and government entities. At the ITSC, a faculty leads the analysis and design of a software solution while students drives the implementation, testing, and maintenance of the software solution. The center currently has 3 full-time professional software developers, 4 faculty members affiliated with the School of Information Technology, and over 20 student interns. The ITSC successfully executes about 20 projects annually  in diverse domains such as criminal justice, retail, data analytics and visualization, education, etc. The rich diversity of domains makes the center attractive to a wide range of students. Furthermore, due to the unique affiliation of the center with the university; the center is able to access the expertise of many faculty in  a wide range of fields within and outside the primary academic unit.

This paper explores the concept of a student-driven software solutions center, examines its key components and characteristics, discusses the benefits and challenges associated with its implementation, and provides information on how we manage operations at the center. The paper also discusses ten main lessons and take-away messages from our experience in managing the center.
By drawing insights from firsthand experiences at the ITSC, this paper aims to inspire and guide educators and institutions interested in establishing their own student-driven software solutions centers.

\spa{
\section{Related Work}
A number of studies have been conducted on student internships, capstone projects, and partnerships between the university and industry. This section reviews the existing works in these three areas. 
\subsection{Capstone Projects}
Bastarrica et al. \cite{bastarrica2017} conducted a survey for students in a capstone project to measure the perceived value of technical and soft kills. Their results show a reduction in the perceived value of technical skills and an increase in the perceived value of soft skills. Vanhanen et al. \cite{vanhanen2012} presented an experience report on a capstone project with industrial customers. Their experience shows that students perceive the course as stressful but extremely rewarding. The students also ranked the course as the one whose content fits best into their specialization. Furthermore, the authors also noted increased participation of external clients. Majanoja and Vasankari \cite{majanoja2018}, Khakurel and Porras \cite{khakurel2020}, and Paasivaara \cite{paasivaara2018does} all reported significant increases in the student's skills and attitude after completing a capstone project with an industrial partner. Perez et al. \cite{perez2012supervision} surveyed 109 capstone project advisors and extracted seven factors of project supervision and seven supervision styles that has been adopted by instructors. Their results show that students tend to score high grades when the capstone projects is focused on technological aspects such as selecting and evaluating relevant technologies.

Sha \cite{sha2021lessons}, Sumner et al. \cite{sumner2008eth} and Tappert et al. \cite{tappert2015real} all shared their experiences in teaching capstone projects. Tappert et al. \cite{tappert2015real} noted that real-world projects with actual customers formed a good learning experience, while Sha \cite{sha2021lessons} mentioned that projects that requires the use of the latest technology often maximizes students' interest. Sumner et al. \cite{sumner2008eth}discovered that when student work on activities that they regarded as fun e.g., game design, they are more likely to learn fundamental technical concepts (e.g., Linear algebra, programming, etc) needed to build such projects. 
Buckley et al. \cite{buckley2004benefits} noted that when students interact with customers who will use the product of the capstone project, the students tends to immediately believe the project is important and they are motivated , they also learn soft skills. However, they often have high expectations and do not want to disappoint the end-users. 
Khmelevsky \cite{khmelevsky2016ten} presented an analysis of 10 years capstone project at their institution. They noted that industry sponsors often request different technology stack from the ones that are taught in school, and student prefer tools and frameworks that are more widely used in industry. The presence of industrial sponsors often serve as a strong motivation for students. Local companies also like capstone projects because they have access to a talent pool where they can recruit to continue to work for them after graduation. However, some student teams suffered from lack of internal leadership and project sponsors may not show up or devote time and resources to the project. Yue et al. \cite{yue2011use} analysed the use of open-source software in 22 real world projects with 14 industrial partners. They reported no consistent difference in mastering open-source software and commercial software. However, the noted that companies prefer open-source software because of its low cost and ease of modification. Students also had an overall positive experience with open-source software.

Many studies have also offer guidelines on assessing capstone projects. Chin et al \cite{chin2017towards} and Farell et al. \cite{farrell2012capstone}, Yousafzai et al. \cite{yousafzai2015unified} provided guidelines for assessing capstone projects. Salem et al \cite{salem2020effective} developed six criteria for evaluating capstone projects. The criteria was used to assess seven projects and their results showed that it was ease to use. Dominguez el al. \cite{dominguez2020factors} discussed assessment factors for capstone projects. They examined 176 capstone projects for discrepancies between advisor and external mentor grades. They noted that student competences have the highest influence on grades by the external committee. They reported that the main reason for the discrepancy is because advisors often capture student competencies during project development while the eternal committee only measures the student competency during the defense. Ahmad et al. \cite{ahmad2011assessment} surveyed 24 instructors in 19 different institutions to understand how they assess capstone projects. They reported that there are variations in assessing capstone projects even withing the same university. They also noted that there is often lack of incentives for faculty to properly supervise capstone projects. Olarte et al. \cite{olarte2015student} noted discrepancies in how student, advisor and evaluation staff assesses student's competencies, project significance, and mentorship received. Both advisor and student often have the same view but student have better perception of their worl and feels they should score higher. 

The approaches discussed so far have focused on capstone projects. Although many capstone projects are executed in collaboration with industrial partners, they still suffer some significant limitations. For example, the priority of students is to satisfy the course requirements and not the long-term success of the projects. Hence, there is no long-term support for the project and the students no longer care about the project after completing the course. Furthermore, a project can be easily abandoned by students if they find it too challenging. The prevalence of project abandonment can significantly affect the university's collaborations with the industrial partners. This paper extends the literature by presenting an experience report on the concept of a student-driven software solutions center where projects are managed by the center and not individual students.


\subsection{University-Industry Partnerships}
Many universities have established partnerships with the industry to provide real life experiences for capstone projects, make it easier for students to secure internships, and provide funding for research activities \cite{pinto2021collaborative}. The industry on the other hand benefits from the university research and access to talent pool \cite{gallego2013knowledge}. Hence, many studies have examined the partnership between the university and industry from diverse perspectives. For example, Pinto et al. \cite{pinto2021collaborative} discusses the alignment of university's research with industry needs, acquiring research funds from industry, and providing real world experience for students. Othman and Omar \cite{othman2012university} surveyed 211 industrial practitioners to understand reasons for sustainable collaboration between industry and academia. Their results show that industry tends to believe they know more than the academia and are likely to question the training regime designed by academia. Hence, industry partners may be unwilling to fund training programs. From the University perspective, a major barrier to a successful university-industry partnership include lack of resources, misalignment with the needs of the industry, unawareness of the latest technologies. Newberg et al also noted that a significant barrier to a successful university-industry partnership are the cultural differences where university want to publicize the research via publications but industry wants to withhold \cite{newberg2001keeping}.

In the same vein, Rybnicek and Konigsgruber \cite{rybnicek2019makes} conducted a systematic literature review of 103 papers to understand the factors that makes a university-industry collaboration successful. They highlighted resource availability, university structure or bureaucracy, and willingness to change as institutional factors that determine success or failure of the collaboration. They also extracted some relationship factors such as clear communication, commitment, trust, and cultural alignment as being important to sustaining such collaboration.  Figueiredo and Ferreira \cite{figueiredo2022more} conducted a systematic literature review on university-industry collaboration from an industrial perspective. Their findings reveal that companies often want to collaborate with academia to drive innovation and acquire talents. This is similar to other findings in the literature \cite{gallego2013knowledge},\cite{fernandez2019funnel}, \cite{ankrah2015universities}. Ankrah et al. \cite{ankrah2015universities} conducted a systematic literature review to understand the motivations, barriers, and outcomes of university-industry collaboration. Their results shows that government policies and incentives often drive the need for collaboration from both industry and academia. Furthermore, partnerships with high mutual benefits are often successful, while bureaucracy, politics, and lack of resources can significantly affect such successful collaborations.

Galan et al. \cite{galan2016drives} conducted a survey with 4,123 responses to understand drivers and inhibitors of university-industry collaboration. They noted that universities are less likely to cooperate if there are no driving motivation even if there are also no significant barrier to such collaboration. They also stated that availability of resources and strong relationships are major drivers to collaboration while lack of funding, cultural differences, and organization structure are barriers. The results reported by Galan et al. \cite{galan2016drives} is consistent with the study conducted by Collier et al. \cite{collier2011enablers}.
Fernandes et al. surveyed 500 firms and concluded that geographic proximity influences cooperation between industry and academia \cite{fernandes2013knowledge}. Deste et al. \cite{d2013shaping} also agreed that geographical proximity makes a collaboration more likely. However, they also noted  companies in technological dense region with complementary skill sets are less dependent on geographical proximity between industry and academia. 
Bruneel et al. \cite{bruneel2010investigating} surveyed 646 participants from 600 organizations to understand barriers to university-industry collaboration. They concluded that lower sense of urgency, mutual lack of understanding, unrealistic expectations, and bureaucracy are major barriers to successful collaboration. Arvanitis et al. \cite{arvanitis2008university} reported that institutions with a stronger orientation to applied research and/or lower teaching obligations are more eager to collaborate with industry. Bstieler \cite{bstieler2017changing} reported that shared governance and similar decision making process builds trust . Goel et al. \cite{goel2017instigates} noted that university employees typically initiates collaborations, while the industry partnes often manage the projects.

Some studies have also introduced techniques for evaluating university-industry pertnerships. Perkmann et al. \cite{perkmann2011should} developed a success map to evaluate university-industry collaborations. The success map can be used to determine the impact of the collaboration based on a set of inputs and outputs variables. Al et al. \cite{al2011balanced} proposed a score card for measuring impact of university-industry collaboration . The score card was evaluated via a survey to 10 firms and the results shows that it is a useful tool to measure, track and improve the impact of conducting collaborative projects with universities.

The studies so far has revealed different motivations, benefits, drivers, and barriers to successful university-industry collaborations. A major discovery is that such collaborations are often initiated by individual faculty or university personnel \cite{goel2017instigates}. Hence, the collaboration may end if the university personnel were to leave the university. Furthermore, the industry partners need to trust the expertise of the university. At the ITSC, the industry partner deals with the center as a unit and not a specific individual. Hence, the successful execution of projects is not dependent on any individual. There has also been a number of institutes affiliated with universities. For example, the  John Hopkins University Applied Physics Laboratory (APL) \cite{banham2017} is a research center affiliated with John Hopkins University and has an annual budget of hundreds of millions of dollars. The APL develops mission-critical systems and technologies for missile defense, naval warfare, and space exploration. The laboratory often deals with safety-critical applications and this may limit opportunities for novice students due to the very low tolerance for error. Another example is the Software Engineering Institute (SEI) at Carnegie Mellon University. The SEI is a federally funded research center affiliated with Carnegie Mellon University. The SEI conducts software engineering research that drives strategic advantage for national security. Similar to APL, the SEI often works in mission-critical and safety-critical systems. Myers et al. \cite{myers2018} presented an experience report on a student-driven information technology support service that  addresses issues related to pharmacy education at the university. Their experience involving 259 cases over 24 months shows that students are motivated and have a desire for engagement when they drive the delivery of solutions. However, the experience report is limited to service management and does not involve an industrial partner. This paper extends the literature to discuss the management of a student-driven and student-led software solutions center with industrial partners.

\subsection{Computing Internships}
Students at the ITSC are employed as paid interns to work on diverse IT projects. This section reviews existing works on computing internship and its impact on student's skillset and employability. Kapoor et al. \cite{kapoor2019} surveyed 40 computer science students who participated in an internship, and the results show that the internship experience promotes personal growth and awareness of industrial expectations. Martincic \cite{martincic2009} presented an experience report on how an internship at an IT solutions company was successfully integrated into the computing department courses as course projects. Their experience shows a significant increase in the students' maturity and technical expertise. Minnes et al. \cite{minnes2021} conducted a survey of 297 students who participated in an internship program. The results show that students focus on technical skills, professional networking, and innate satisfaction that their project will be used in the real world.

Groeneveld et al \cite{groeneveld2019} noted that internship and capstone projects aid students to recognize the significance of developing essential non-technical skills such as collaboration and conflict management required for successfully executing software projects. Dean et al . \cite{dean2011} noted that the benefits obtained by a student in internship or capstone projects depend on the role of the students in executing the projects. They also noted that an unpaid (volunteer position), real-world, student-driven, and long-term project brings significant and valuable skill sets to the students. They noted that the greatest advantage of projects with such characteristics is the long-term nature of the project because they motivate students and allow learning to be the primary focus of the students when executing the projects. Contrastly, Kapoor et al. \cite{kapoor2020} noted that there were no significant academic differences between students who participate in internships and those who do not.  Jaime et al \cite{jaime2019effect} noted that participating in internships prior to capstone projects improves students' technology, methodology, and project management skill set. They also noticed an increase in the complexity and technological novelty of the resulting capstone projects, and reduced advisor involvement in the practical project execution. However, they also noticed an increase with advisor involvement in monitoring student work (such as meetings, reports, and initial arrangements). Sudol and Jaspan \cite{sudol2010analyzing} showed that internships tend to reduce misconceptions about software engineering. 

Change et al. \cite{chang2016relationship} conducted a survey with 667 participants from 32 universities to understand relationship between students background and internship conditions. Their results show no significant relationship between internship performance and academic performance. Furthermore, higher year student (e.g., final year students) have higher participation rate in internships.
Lehman et al. \cite{lehman2024sealing} surveyed 1,018 students from 15 US universities to investigates how internship experiences affect computing careers. Their results shows that students often felt that internship experiences increased their professional (or soft) skills. Furthermore, women are more likely to report increase in technical skills. They also noted that growth in skillset and strong self-efficacy are the strongest predictors of increased interest in computing career. Hence, they conclude that internships play a critical role in retaining undergraduate women and non-binary students in computing when transition from college to career.
Stepanova et al. \cite{stepanova2021hiring} survey of 218 recruiters. They reported that software developer is most common job and recruiters often emphasize technical and coding tests during the recruiting process. They also reported that work experience, GPA, and project sections are the most significant part of a resume of CV that influences hiring choices.
Kapoor et al. \cite{kapoor2020barriers} surveyed 320 students from two US universities who did not participate in an internship to understand barriers to securing internship. They discovered four main themes - low self-efficacy, actions (lack of preparation or reliance on course work), alternate priority, and application process challenges. Their result was consistent with Wolf et al. \cite{wolf2023modeling} who surveyed 518 students.

Ntafos and Hasenhuttl\cite{ntafos2015internships} examined the work history of 225 students to understand the correlation between the student's  workload and academic performance. They reported that internships  helps students support the cost of their education, gain practical experience, self-confidence, self-esteem and significantly improve their prospects in the job market. Internships also motivate students to complete their degree thus it acts as a means to improve retention and graduation rates. However,they also reported that some students may struggle to balance work and study and performed badly in both.
Kang and Girouard \cite{kang2022impact} analysed internship reports of 42 graduate students over 6 years to understand the impact of internships in User Experience. Their results show that students gain technical skills, learn more about industry culture, soft kills, empathy towards end users, and realization of future career path. Conrad \cite{conrad2021colleges} analyzed student experiences after completing their internship. The students believe that practical application of skills, continuous career development, and best practices on the job are critical to a successful internship. They recommend promoting internship as early as possible and providing a class to prepare them.
Olarte et al. \cite{olarte2019impact} also reported that part-time internships improved quality of capstone projects. They noted that a major benefit with part time internship is that students can continue to maintain contact with the university. However, the workload and multiple course coordination are significant challenges for students. Mia et al. reported \cite{mia2020conceptual} potential problems with internship. They include difficulty in placement of interns to the industries, gaining real-life experience, managing learning process during the period of internship, and measuring the learning outcome during and after an internship.



Saltz and Oh \cite{saltz2012open} discussed an immersion experience partnership between universities and academia where students participate in hand-on work-based learning experience that augments the classroom experience. They developed a coop program that complements typical CS curricula with large scale enterprise applications.  They surveyed 31 students to evaluate the program and the results show that there was significant improvement in students soft and technical skills. The most significant increase in competency was determined to be in understanding work place corporate culture. 
Menezes et al. \cite{menezes2022open} presented an experience report where experienced software engineers volunteer to mentor students who work on open source software. They reported that students who received stipends were more likely to commit 40 weekly hours  to the project. Furthermore, 69\% of the participants secured a job within 12 months of graduation and they are 12\% more likely to secure a job. They also noted significant improvement in technical skills, ability to build software independently, and enhanced collaboration skills to work with peers.
Sweetser et al. \cite{sweetser2020setting} reported on an internship course model where students are provided three streams options; internship with a company, academic project based program, professional mentoring workshops. Massi et al. \cite{massi2016transforming} investigated how social structure affects how students identify with engineering. They created a social learning community involving interns, industry professionals, faculty, and support staff. They then incorporated value-added program activities including social events, distinguished Speaker Series, and an annual Symposium. The authors reported that social activities and relationship-building experiences often had the most impact on women and first generation college students because they create a welcoming environment. Hence, they concluded that participation in any of the social events has a significant impact on commitment to the major, building confidence, and feeling welcomed into the community.





The approaches discussed so far have focused on software internships. However, the student work environment is often distinct from the university during an internship. This distinction makes it challenging to monitor and evaluate a student's progress. Furthermore, many companies do not trust interns therefore, students do not often drive the project execution. This paper extends the literature by presenting an experience report on the concept of a university-owned and student-driven software solutions center.
}





\spa{
\section{Methodology}
The main goal of this paper is provide a blueprint for other institutions to establish a similar student-driven software solutions center for enhancing student skills with real world experience. To achieve this goal, we developed the following research questions.
\begin{enumerate}
    \item [RQ1:] Why do we need to establish a student-driven solutions center? A strong rationale for a solution center would serve as a needed motivation and catalyst to develop a similar center. This is because a student-driven solution center would normally require dedicated time and resources during its inception. Hence, it is necessary to enumerate its importance to ensure that the needed resources are provided. This research question enumerates the need for a student-driven solution center using the ITSC as a case study.
    \item[RQ2:] How to sustain a student-driven solutions center within a university setting? The successful establishment of a solution center is a great achievement \cite{oster2019improving}. However, such center is likely to fail if proper processes and measures are not implemented. This second research question seeks to investigate how the ITSC was sustained and managed over the last twelve years. The goal is to provide a blueprint for other institutions.
\end{enumerate}
To answer these research questions, we adopted a case study approach using the guidelines by Yin \cite{yin2018case}. Specifically, we analysed archival records at the center and conducted critical personal reflections from the authors who had been at the center from March 2012 till date (March 2024).
\subsection{Data Collection}
The first step is the data collection process where we reviewed the 12-year data that has been documented at the ITSC. The data sources include intern application forms, project contracts, email communications with clients, git repositories, and slides of presentations given on behalf of the center. While majority of these data sources are confidential and cannot be shared publicly, we provided an aggregate overview of the extracted data to answer the research questions. Figure \ref{interns} shows the number of students that has successfully interned at the ITSC each semester (both new and returning interns). 

\begin{figure*}[htbp]
\centerline{\includegraphics[width=\textwidth]{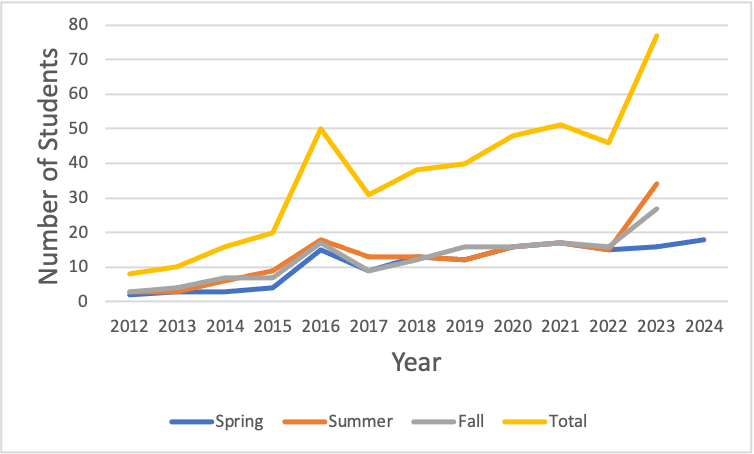}}
\caption{Number of Student Interns per Semester}
\label{interns}
\end{figure*}
\subsection{Brainstorming Activities}
After the initial analysis of the archival data, we then set up a series of brainstorming activities among the authors. The first author had observed activities at the ITSC for two semesters to understand the day-to-day operations of the center. The first author also served as an independent evaluator and moderator for the discussions during the brainstorming activities. The second author joined the center in 2014 and rose through the ranks at the center. She has served as a student intern, co-op student, graduate assistant, research associate, and she is currently a faculty-fellow with the center. The third author is the founder and current director of the center. The brainstorming activity was used to gain further understanding of the archival data, and provide a historical perspective based on the personal observation of the center over the years. 

A student-driven solution centers revolves around three main objects- the students, clients, and projects. Therefore, the brainstorming activities was conducted to unravel the reasons behind some major decisions during the ITSC evolution, provide more insights to the archival data, and answer the two research questions with respect to the students, clients, and projects. The following questions were used as a guide for the discussions.
\begin{enumerate}
    \item How did it all begin and why do we need to establish the center?
    \item What made you join the center as a student and what was the first project you worked on?
    \item Can you compare the time you were a student and a trainee with now that you are a faculty fellow and training students?
    \item How are students employed and engaged at the center?
    \item What factors do you think empowered the students?
    \item How are the projects managed?
    \item How was the center sustained over the years?
    \item How do we manage client's expectations and build credibility with the client?
    \item How did you manage center finance including revenue, salary negotiations, and budget?
    \item How do you determine the stipend to be paid to the intern?
    \item What are the challenges in managing the center and what are the opportunities for growth?
    \item How did we navigate obstacles and manage outcomes?

\end{enumerate}


These discussions were transcribed and analysed to help answer the research questions. Section \ref{sectionNeed} and Section \ref{sectionManage} present the results of the analysis and answers to the research questions. 
}
\section{ITSC- A Student-Driven Software Solutions Center}\label{sectionNeed}
The Information Technology Solutions Center (ITSC) was established on March 2012 shortly after the Department of Information Technology (now called the School of Information Technology (SOIT)) was created at the University of Cincinnati. The ITSC was established by the third author, who is also the current director of the SOIT. The ITSC is a unit under the SOIT, and it is subject to the rules and regulations of the University of Cincinnati. Henceforth in this paper, "the center" refers to ITSC while "the school" refers to SOIT. 

\spa{The founder of the center identified a need for the students to engage in real-world projects, while also identifying the industry need for low-cost software solutions. These intersection of these needs actually led to the conception of a student-driven solution center. Furthermore, }at that time when the ITSC was created, the SOIT offers only bachelor's degree with few faculty and students. The SOIT was a teaching-only unit and faculty were not expected to conduct research. However, the faculty knew that the school will need to engage in edge-cutting research to survive and grow within the university. The ITSC was formally established to drive the engagement of the school with industrial partners. Specifically, the following reasons drove the establishment of the ITSC. 
\begin{enumerate}
    \item Name Recognition and Building Credibility for the School. As a newly-formed academic unit, it was important for the school to be recognized to attract leaders in the practice. Having a strong collaboration with the leading players in the industry provides an avenue to demonstrate the worth and significance of the school. This also ensures the students can easily get job placement with these industries after graduation.  
    \item External Funding. The projects executed by the center are funded by industrial clients. This provides a source of revenue that can be used to drive other initiatives in the school. The external funding also funds the students that worked on the projects.
    \item Enable the growth of research programs. Having a rich source of funding is important to drive the growth and quality of research executed by faculty members. However, another important driver for research growth is relevant problems that research should address. The ITSC exposes the faculty to real-world problems that their research can address. This provides an avenue to ensure that the impact of the research is significant. The ITSC serves as a source of external funding and a means of identifying practical research problems, thereby driving the research growth of the school.
    \item Distinguish the students with industrial projects. The industrial experience provided by the ITSC to the students also provides an opportunity for the students to distinguish themselves from students at other universities or programs. Many students at the center often lacked confidence and have little technical expertise when they first joined the center. The center provides an opportunity for the students to grow by enhancing their hands-on technical skills, problem-solving skills, and communication skills. By the time the students complete the third or fourth year of their bachelor programs, most students are well equipped and skilled to make a significant contribution in many aspects of Information Technology.
    \item Generate Income for students. The center also provides an avenue for students to earn income while working in a space related to their field of study. This is often rare to find for most students. The center is also flexible with the student schedules and can accommodate student-centric demands such as creating a new university bus route to the center. Furthermore, international students in the United States cannot normally work outside the university. The center also provides an opportunity for these groups of students to work in a related field without jeopardizing their legal immigration status.
    \item Provide opportunities for students to innovate and become independent thinkers. The student-driven nature of the center provides students with a sense of ownership when working on projects. This allows them to innovate and provide creative solutions that satisfy clients' requirements.
    \item Enhance faculty personal development. The center also provides a great opportunity for the professional development of faculty and staff in the school. The third author has a PhD in Aerospace Engineering and initially did not have enough skills and credibility in building software solutions. The center provides a platform for the second author to build skills and demonstrate credibility in building software solutions.
    \item Increased income for affiliated faculty. The ITSC also provides  opportunities for faculty to collaborate with the center and earn extra income.
\end{enumerate}

The ITSC was able to succeed because it fills a specific need for industries which is the execution of projects that are important to have some budget, but not a priority to the organization. The cost of executing such software projects is significantly minimized when the companies outsource the project to us. The center was also able to clearly articulate this need to the companies, and this convinced the companies to outsource projects to us, even when they knew the project execution will be driven by students. The majority of the early projects executed by the center came because the companies have an idea they wanted to explore, but nobody wanted to allocate significant resources to execute them.

Another reason for the success of the ITSC is that extensive time and effort was spent on analyzing, refining, and formalizing the client's requirements. The majority of the clients often have an idea about what they want, but they could not clearly articulate them. The center was able to utilize the domain expertise of the diverse faculty in the school to help with the requirement analysis of client needs. For example, a client came to the center after considering many other options because the center was the only one that clearly understood what they wanted, even though the center was not the cheapest option.

In summary, the establishment of the center provides considerable benefits and opportunities to students, faculty, and the School of Information Technology. The center also provides a low-cost option for industrial partners and provides a pathway for industries to access the research expertise of the faculty. The self-sustenance of the center for over a decade is proof of the center's significance in providing value to students, faculty, the university, and the industry.

\section{Managing the Center Operations}\label{sectionManage}
The normal operations of the center involve the clients that provide projects to the center and the students execute them. This section discusses three main items related to managing operations in the center and they are student management, project management, and client management.  Student management deals with how to recruit students, and how to train and support them to gain the necessary technical expertise for the workforce. Project management deals with the execution of projects with respect to the contract signed with the client. Client management deals with managing the expectations of the client so that the learning curve of the students does not negatively impact the client's confidence in the center's ability to successfully complete the project and also ensure that the client feels that the cost paid for the project execution is worth it. 
The section also discusses how we have sustained the center over the years \spa{and the software engineering methodologies that has been adopted}.
\subsection{Student Management}
The student is at the core of the center and the efficient management of the students is crucial to the success of the center. It is important to select the right student and provide continuous opportunities to engage the student.

\subsubsection{Student Recruitment and Resources}
The University of Cincinnati's Cooperative education (co-op) initiative \cite{coop} often requires students to work full-time in alternate semesters or part-time during regular semesters, in a workplace related to their field of study. This provides opportunities for students to gain industrial experience and ensure they are adequately prepared to address the needs of a transforming workforce. A major challenge for many students is finding internship opportunities in a related workplace \cite{kapoor2020barriers}. The center provides an opportunity for students in the coop program to work in a related field of study. The center also accepts non-coop students part-time (during regular semesters) or full-time (during the summer) in the center. Hence, there is usually a large pool of students the center can recruit every semester.

Recruitment of the right student is important to ensure the continuous success of the center. We prioritize a student with the right ethical standards and basic technical knowledge. There is often an advertisement that is sent to the school every semester to recruit new students. Then the students are screened and interviewed based on the current needs of the center. The center occupies an entire floor in an 8-storey university building. This provides an open space that is large enough to fit many students. The center provides a workstation and an enterprise GitHub account for each student. 

\subsubsection{Student Onboarding}
\spa{The onboarding process is a way for new employees to acquire the necessary skills and information to become a successful employee \cite{cable2013reinventing},\cite{nalband2017employee}. The onboarding process is a very important part of any organization since it directly impact retention rate and employee self efficacy\cite{gupta2018relationship}. The onboarding process at the center has been set up to address the common challenges of integrating new employees \cite{chen2010suggestions}. The onboarding process uses a combination of internal documentation, code review, induction, team leader support, peer support and self learning elucidated by Buchan et al. \cite{buchan2019effective}. The entire onboarding process is described in the paragraph below.}

When a student is selected, there is a set of activities aimed to familiarize the student with the center. Students are onboarded through a series of presentations generally initiated by a full-time staff/faculty member. These presentations orient the students to the work culture of the center, expectations of the center, initial system setup, system walkthroughs, and knowledge transfer from existing students or staff. Recently, we introduced a starter project for new students to measure their technical expertise and gauge their readiness to work on live projects. The starter projects require new recruits to build an application that can manage the resumes of applicants for a position. The starter project allows us to identify the problem-solving skills of the student and the resume project also encompass all the technical skills necessary to execute the current projects at the center. Hence, the resume project has to be developed as a multi-tier application, use the technology stack that is being used at the center, and include features such as file management, user roles, session, authentication, etc. These features are common to most of the applications that are built by the center. \spa{Students are assigned to teams based on their interests, technical expertise, project needs, and performance on the starter project.}
\subsubsection{Student Engagement}
Students in traditional internships are often assigned to a team and asked to work on the team project when they resume. This can be intimidating for students with a longer learning curve. At the center, we also assign students to a team. However, each student is often allocated a specific independent task within the team. This is structured in a way that other team members are not dependent on the completion of the student's tasks but the student's work can later be merged with the team's efforts. This approach communicates to the student that we trust the student to deliver the task. However, we ensure that only the execution of the project is driven by the student. The communication with the client is handled by the leadership team at the center. The center also ensures there are regular meetings with the students. Every team has a team channel and holds regular stand-up and sprint meetings. The center also has trained and certified scum professionals that help with adapting the Scrum processes and methodologies\cite{srivastava2017scrum} to suit the center's operation.

The center continues to attract students by offering competitive pay and flexible schedules around the student classes. The center provides opportunities for students to join on a part-time basis during the Spring/Fall semester or a full-time basis during the Summer semester. The students can also work on a full-time basis during the semester they are on coop at the University. The center is under an academic unit; therefore, it understands the needs of the students and can provide appropriate accommodations than many companies.  Hence, more students opt to join and remain at the center for the duration of their program. Many companies train student interns by giving them explicit and defined tasks to execute. The students are not in charge of the projects and do not feel a sense of ownership. On the other hand, the center provides students opportunities to drive the execution of projects, attend meetings with the clients, and explore creative solutions to address challenges. The students stay with the center for about 12 months on average, while a good portion of the students stays for about the market average of 24 months \footnote{https://www.zippia.com/software-engineer-jobs/demographics/}. Furthermore, the faculty also teach classes in the school and they can identify students who would be a good fit for the center.

\subsection{Project Management}\label{projectmgmt}


The successful and timely completion of the project is essential for boosting students' confidence and satisfying the client's expectations. Most of the projects executed by the center came through word-of-mouth and personal recommendations. The initial set of projects often generates little revenue and is unable to sustain the center. However, the successful execution of the small projects allow us to propose even more ambitious features with a larger cost to the clients, and many of the clients accepted the proposed extensions. For example, the initial project with a Fortune 500 company in the retail sector, started with a small amount that has now grown to a significant annual contract that is worth more than five times the initial contract. The contract has also been renewed continuously every year for the past six years. This demonstrates the client's satisfaction with the center.

Many of the initial clients that came to us were restricted by their budget and often had no option but to work with us. Hence, there was no reason to educate the clients on why we believed students can handle the projects. The successful execution of the projects has enhanced the credibility of the center and the confidence of the clients to trust us with even more challenging projects. The center also adheres to standard values in software development by focusing on value creation, iterative and agile development, and constant engagement with clients. The students are also involved in every meeting with the clients and a working prototype of the product is usually ready within one month.

The center adopts an iterative project management \cite{bittner2006managing} to create rapid prototypes and validate the client's project expectations \cite{mchugh2012impact}.  This enables to achieve early demos or a working prototype that can be presented early to the client. The iterative approach also allows us to easily adjust to evolving vision of the client and manage any change in the scope of the project. The center also ensures that the execution of projects adheres to stand project management practices \cite{crawford2007global}. Specifically, the center ensures that standard project management standards are followed to manage the scope of a project, manage schedules with weekly and monthly deliverables, and  develop a communication plan with the client to ensure that the clients are carried along throughout the project life cycle.



\spa{
\subsection{Software Engineering Processes}
The center has adopted some standard software engineering process over the years in order to achieve its objectives. This section discusses some of the software engineering processes that are currently being implemented in the center.
\begin{enumerate}
    \item \textbf{Agile Methodology}. Section \ref{projectmgmt} mentions that the center adopted an iterative project management approach. Hence, in order to execute the iterative project management process, the center uses the agile methodology\cite{beck2001manifesto} and scrum process \cite{srivastava2017scrum} throughout the software development process. This allows the development of rapid prototypes that can be presented to clients as soon as possible.
    \item \textbf{Requirement Elucidation.} The requirement elucidation phase is very important to the success of the center because it is the first step in ensuring that the final product meets the client expectations. Hence, the center adopted a client-driven approach in which the customer success and requirements are prioritized \cite{racheva2010conceptual}. The requirement elucidation process usually starts with an initial overview brief from the clients, followed by a brainstorming section with the client. The final, detailed set of requirements are prepared by the center and signed off by the client. The standard elucidation techniques such as interviews, scenarios, and prototypes are used based on the nature of the project \cite{sutcliffe2013requirements}. The center also uses a quality-function-deployment model to ensure customer satisfaction \cite{bouchereau2000methods},\cite{ginting2020product}. The interns attend client meetings and participate actively in the requirements elucidation process.
    \item \textbf{Design.} The center uses a loose UML approach \cite{chaudron2012effective} for analysing and constructing the design of the system. The main goal of the design phase is to analyse and understand the system, hence there are no strict adherence to majority of the UML rules. Furthermore, many of the design decisions are left open during the design phase. This provide opportunities for students to explore creative solutions during the implementation phase.
    \item \textbf{Implementation and Testing}. The center does not have a standard programming methodology for all projects. The programming paradigm that will be adopted for the project will depend on the nature of the project, programming language to be used, and student expertise. However, every intern is expected to write unit tests to validate the correctness of all methods and classes they create \cite{khorikov2020unit}. This is also another team of testers who ensured that the final product satisfies the clients' requirements.
    \item \textbf{Maintenance} All the projects executed by the center are developed on a secured enterprise git repository for easy collaboration and automated tracking of changes to the software artifacts \cite{spinellis2012git}. However, majority of the projects executed by the center are one-of projects that are handed to the user after its completion. For these one-of projects, the center is not involved in the maintenance operations. However, there are some projects that are currently maintained by the center and needs to adapt to changing client requirements. For such projects, the center is focus on continuously providing value to the clients by first understanding the clients mission and what it needs to succeed, proposing new features that can help fulfil the client's mission, and continuous engagement with the client to maintain client satisfaction. 
\end{enumerate}
The student interns participated actively in all the phases of the software development process and they actively implement the software engineering processes discussed above. The only exception is during the client meeting where the students only observe or respond to questions and do not actively contribute to the discussions. The is due to the sensitive nature of some of the discussions and the naivety of the interns because majority of them are novices. However, the students are active in other phases of the requirements elucidation process and the the rest of the software development process.
}
\subsection{Client Management}
The satisfaction of the client is crucial to the sustainability of the center. A critical part of managing a client is to understand the problem the client is trying to solve and then use the expertise of the faculty to articulate the problem. Hence, the main focus of the client management process at the center is to make the clients feel that the center understands the client and the problem, instead of trying to prove that the center has the right technical skills. The center also carefully selects the clients to work with due to the inherent risk involved in a student-driven execution of projects. The center communicates early and regularly with the clients to manage their expectations.

 The center allows the client's vision to evolve without incurring extra cost and the client is not limited by the scope defined in the initial contract. The center adopts a fixed-cost model that allows the client to modify the project's scope and still deliver on time. The center follows an iterative process to execute the project, schedule regular meetings with the clients, understand and continuously keep an eye on the problem the clients are trying to solve, and identify critical features and wish lists for a software solution that addresses the problem. The features on the wish lists may be implemented during the next evolution or release of the project. Any major decisions on the evolution of a project are made jointly between the center and the client.
 
 
 The center is focused on providing value to the client. The center also builds credibility by taking responsibility for any shortcomings, acknowledging mistakes, and responding quickly to client requests or messages. This approach has continuously led to contract renewals and long-term partnerships with clients. Furthermore, the center often treats scope creep as delayed learning; thereby allowing the center to learn from its mistakes and adjust to client demands.
 
\subsection{Sustaining the Center}

The center is a unit under the University of Cincinnati, therefore it has to conform with the university regulations on contracts and finances. There is also a need to figure out how to estimate the cost of a project and charge clients. The charges by many companies are not fixed and vary based on the dynamic time and effort spent on the project. However, the center adopted a fixed-cost approach for each project, where the cost of a project is determined before the project execution. This comes with an obvious risk of undervaluing the project and losing money. However, the fixed-cost approach endears companies to the center and allows the companies to continuously patronize the center.

The center has also prioritized paying competitive stipends to the students working on software projects. Through data collected from the Coop office at the University of Cincinnati, the center was able to determine the average stipend for student interns and provide pay that is above the average. The students also get a pay increase for every semester they work with the center.

Despite lots of planning and careful adherence to industry standards, a lot of issues can arise from the student or client-side during the execution of a project. Hence, to navigate any obstacles we often adhere to the following principles.
\begin{enumerate}
    \item \textit{keep an eye on the goal}. The main goal of the center is to provide technical experience for the students and generate revenue for the center. The student's confidence is often tied to the successful completion of the project and the successful completion of a project that satisfies the client often increases the revenue and reputation of the center. A negative review from a bad client can adversely affect the center. The weekly and monthly meetings within a project team and with the client are geared towards achieving this goal.
    \item \textit{Shuttle diplomacy}. When a client is not satisfied with the progress of a project, the center strives to create an in-person meeting either at the client's venue or at the center. The center may also create a mediation panel comprising key figures within the university to meet with the client and reach an amicable resolution.
    \item \textit{Student ownership.} The student is entrusted with the project's execution and the center works to make sure the student has a sense of ownership in the project. Naturally, this motivates the students to work towards the successful completion of the project and take responsibility for any setback.
    
\end{enumerate}


The center follows standard practices that are obtainable in the software industry including iterative project development, scrum and agile guidelines, project management standards, etc. However, many of these standards are geared towards industries with professionals and not students. Hence we adapted the standards to suit our student-driven model rather than modifying our model to conform with the standards. This allows us to not overburden the student and remove some bureaucracies from the student.

\section{Sample Projects Executed}
The projects of the ITSC range from projects within the university's different organization units to public and private organizations. The ITSC clients span a wide variety of sectors including but not limited to Criminal Justice, Retail, Cybersecurity, Healthcare, Human Resources and Professional Development, and Manufacturing. Some of the projects are highlighted below:
\begin{enumerate}
    \item A Suite of Applications for Criminal Justice: This suite of web applications has been in production for seven years. This system was built in collaboration with the University of Cincinnati Corrections Institute (UCCI) to provide a user-friendly platform to aid officers in performing their duties. The tools in the system are based on UCCI research. Initially built for a county in the state of Ohio, the system has since been implemented state-wide in multiple states as well as smaller counties and agencies.

    
    \item Innovative solutions for consistent interpretations within the legal domain: This solution provides a platform that allows users to reach a consistent interpretation of the law based on previous interpretations of similar scenarios. The platform includes an informative website  and a labelling platform. The project is part of the suite of projects to support a government agency in Ohio with some of its technology needs.

    \item Fortune 500 Organization Web Application: This project started when an IT manager in the organization unit had an idea to optimize their processes. The ITSC was able to address the business need by engaging in a pilot project to convert a Lotus Notes application to a more modern technology stack. Based on the success of this, the organization's projects became an annual contract and the partnership has been longstanding for seven years. 

    \item Virtual Events: During the pandemic, the School of Information Technology at the University was faced with the unique challenge of taking a 1000-person annual event called the IT Expo online. The IT Expo is an opportunity for graduating seniors in the program to showcase their work to the community and win awards. The event also includes a high school competition and a research symposium for the graduate program. The ITSC conceptualized and executed an online platform to live stream videos, host posters, and help the community engage with the projects through an online discussion board. Based on the success of the event, other units in the University also contracted the ITSC to host their annual events on the platform. Ranging from symposiums and research talks to award ceremonies, the ITSC hosted several virtual events. These were eventually discontinued due to their time-critical nature.

\end{enumerate}

The ITSC has a niche in taking projects that are critical enough to have a priority, but not critical enough to have a high budget in an organization's technical arm. The projects taken on are not time or life-critical, which allows space for students to learn and grow.


\section{Opportunities for Growth}
The center has grown very much over the last 12 years. However, we believe there are still important avenues for continuous growth and improvements. This section discusses areas and directions for growth that has not yet been explored by the center. This vision may also assist any computing department that wants to establish a similar center on potential opportunities for growth. The possible areas for growth are listed below.
\begin{enumerate}
    \item Expansion to high school. Another vision of the center is to establish branches in high schools and provide mentorship for the staff and students at the high school on how to successfully manage the day-to-day operations. A number of efforts have been allocated toward expanding computing education to K-12. We believe an expansion of the center's operations to high schools will aid in realizing the goals and objectives of K-12 computing education. Furthermore, the School of Information Technology already has an Early IT program \cite{earlyit} where high school students can complete first-year college courses while in high school. The center can extend the Early IT program to support student-driven industrial projects.
    \item Become a leading player in a specific domain. The center accepts software projects in diverse domains and we believe that we can explore University resources and faculty expertise to provide software solutions in any domain. However, about 80\% of our current projects are in criminal justice and we have gained a lot of experience in the social sciences over the last decade. Hence, an organically evolving vision of the center is to become a leading software solution provider and innovator in the social sciences. Furthermore, having deep expertise in a specific domain also enhances the credibility and visibility of the center. This should also motivate industrial customers to engage the center in addressing critical problems in the social science domain.
    \item Increased engagement with students for the undergraduate capstone project and graduate student project. Currently, we mostly provide opportunities for students to join the center as interns. There are limited opportunities for students to complete their capstone project with the center. However, we believe that creating opportunities for such students to collaborate with the center for their capstone projects will provide the students with a unique opportunity to work on a real-world project that is in production.
    \item Ability to scale and diversify to numerous domains. The University of Cincinnati is an R1 university with very high research activity. The university currently has over 48,000 students and 6,000 faculty in diverse domains and fields of expertise.  The center can scale as fast as possible due to the huge pool of available student numbers. The center can also collaborate with faculty in different academic units within the university to provide innovative software solutions in diverse domains. 
\end{enumerate}
This section has discussed some opportunities for growth at the center over the next decade. The opportunities enumerated above are non-exclusive, but we believe that they may provide extra motivation for colleges and departments willing to establish a similar center.

\section{Challenges}
The center has encountered a number of challenges that limits the impact or smooth operation of the center. These challenges are discussed below to prepare prospective colleges for potential setbacks they need to prepare for in order to develop and sustain a successful student-driven solution center. These challenges are ongoing and the center continuously tries to address them. The challenges include the following.
\begin{enumerate}
    \item Moving from \spa{testing environment to production environment}. When working on building a new software solution from scratch, students are better empowered because they have more freedom to explore and they can be creative. Furthermore, the students tend to work on a lot of design problems.
    Inevitably, the successful completion of projects means the project moves to a production environment where the software solution is live and currently in use. A live project means there is little room for mistakes and such projects might be critical to the operations of the client companies. Hence, the process in production systems needs to be standardized, thereby limiting opportunities for students' creativity. Furthermore, projects already in production environment also require more rigorous onboarding and vetting of students, strenuous training, and a limited margin for error. 
    Unfortunately, the center has seen an increased number of systems in production environment that we currently maintain and upgrade due to the successful execution of past development projects. For example, one of the projects currently has over 10,000 users. 
    \item Opportunities for students with zero-technical experience and skills. The main strength of the center is that we have been able to provide opportunities for students with zero-technical experience and skills to join the center, learn valuable skills, and be empowered in the center. These are the students that nobody will likely hire, but they succeeded in the solution center. However, with the increasing number of time-critical and mission-critical projects that are being managed by the center, it is increasingly difficult to recruit students with zero skills to work on such critical projects. For example, recent research assistant positions often require experience. 
    \item Providing continuous opportunities for long-term students to grow within the center. Students are initially assigned to an entry intern role when they joined the center. The students are then assigned to a project team. Over time, the student might eventually become the team lead and serve as a mentor for new interns. However, the roles are limited at higher levels, and students might be stuck at a specific level for a long time. Hence, there is the challenge to always create new roles and challenges for long-term students to keep them engaged and motivated to stay at the center.
    \item Managing student transitions between semesters. Some students might choose to leave the center at the middle of the semester while some students might not renew their appointments at the end of a semester. This creates a vacuum in completing the projects that the students were handling before separating from the center. Another challenge is how to retain the knowledge and experience the students have gathered while working at the center to ensure that new incoming interns do not have to repeat the same mistakes and perform well in the position. \spa{Frequent tunrover can negatively affect employee morale due to time spent from the task to train new employer \cite{bhattacharya2015employee},\cite{pike2014new}. One way to mitigate this is to have an overlapping period where an incoming intern works with an outgoing intern for some time.}
    \item Challenges with growth. As the center continues to grow and the number of students increases, a number of challenges such as limited team bonding and forming of cliques have emerged. These challenges emerged organically due to the increased number of student interns and the challenges may continue to escalate higher if it is not mitigated. A number of measures such as increased social activities aimed at team bonding, anonymous feedback from students to track discontentment and isolation, and regular meetings have been implemented at the center to mitigate some of these challenges. Hence, it is imperative for colleges willing to establish similar centers to watch out for these social concerns.
    \item Continuous training for the leadership team. The center is currently managed by a 7-member leadership team. Sometimes, a member of the leadership team might leave the center or a new member joins the leadership team. Furthermore, new standards, protocols, or processes might need to be established in order to address evolving challenges or foster the growth of the center. It is also possible that the newly established process might not be geared toward student empowerment. Hence, there is a continuous need for sustained professional development training for the leadership team on how to empower the students. There is also a need to continuously emphasize the student-centric nature of the center, and ensure that all the operations in the center are directed towards student empowerment.
\end{enumerate}
The challenges discussed above are aimed to prepare potential creators of student-driven solutions centers on the possible issues they may encounter. This should make them more prepared and ensure a successful solutions center. \spa{A number of plan has been initiated to mitigate these challenges. For example, one of the current plan for the center is to source for grants and initiate in-house projects where students with zero-experience can be trained and prepared for critical projects. This will also create opportunities for long-term student to become team leads and create a path way for growth. Appropriate measures will be put in place to ensure that such projects are up to the required industry standard. }

Finally, we at the center believe these challenges also present research opportunities whose impact goes beyond the solution center. For example, an effective solution for managing student transitions might also be adapted by technology companies to help mitigate the negative outcomes associated with a key developer suddenly leaving the company in the middle of a project's execution.


\section{Discussion: Lesson learned and take-away messages}
In this section, we discussed the lessons learned and take-away messages from our experience in developing and sustaining a student-driven software solutions center. \spa{These lessons were extracted during a brainstorming session among the authors. 
}
\paragraph{\textbf{Lesson learned 1.} A university-affiliated, student-driven solutions center provides a unique set of opportunities for students, faculty, information technology or software engineering programs, and industry} The center allows students to gain important industrial experience and drive the development of a real-world product. Student satisfaction is also evident in the high retention rate and increasing number of interns over the years. For industries, it provides a low-cost but efficient opportunity to execute important but non-critical projects or ideas. Many of the clients have been with the solution center for years and the revenue from projects has kept on increasing year-over-year. This demonstrates the satisfaction of the industrial clients with the project's execution. For faculty, this presents an opportunity to conduct research and expand their network with industrial clients. The center increases a program's credibility and ensures that the students are adequately prepared to meet the demands of the workforce upon graduation.
\paragraph{\textbf{Lesson learned 2.} With proper mentorship, students can be trusted to drive the development of software solutions} The center has executed dozens of software projects. The execution of these projects has all been led by students under the guidance of a full-time developer. The full-time developer mainly provides mentorship and handles the transition of students across each semester. However, the design and development of the software solution have always been driven by the students. The successful execution of all the projects and the satisfaction of the clients shows that students can be trusted to build industrial software solutions.
\paragraph{\textbf{Lesson learned 3.} Tolerance for delays is important} When working with students, there may be a learning curve that can delay the execution of the projects. Hence, it is important that a similar center should only deal with clients and projects that have a level of tolerance for delays in the planned project timeline. However, it should be emphasized that the quality of the solutions will not be compromised.
\paragraph{\textbf{Lesson learned 4.} Efficient client management is key to the success of the center} In any business environment, the client is the key to success. The satisfaction of a client is important for the growth, sustainability, and revenue generation of the center. An unsatisfied client may negatively affect the reputation of the center. Hence, it is important that communication with the client is handled directly by the management and students do not have to communicate directly with the client. However, students should be invited to client sessions as an observer and not as an active participants. Furthermore, it may be necessary that the management team is trained in efficient client management techniques including communication and conflict resolution.
\paragraph{\textbf{Lesson learned 5.} Not all potential clients or projects should be considered} Every possible client or project is a potential source of revenue. However, due to the student-driven nature of the center, it is important to avoid safety-critical or time-sensitive projects. It is also important to communicate these clearly to the clients and ensure that realistic expectations are set. Hence, not all potential clients or projects should be considered by a student-driven solution center. Incompatible projects will inevitably lead to client dissatisfaction and affect the center's reputation. 
\paragraph{\textbf{Lesson learned 6.} Efficient analysis of the client's requirements is important towards client satisfaction} Majority of the clients continue to renew their contracts with the center because the center understands the requirements of the client's needs. Efficient requirement analysis skills are important and crucial to client satisfaction. The center always strives to understand the client, the role they play in their organization, and how the successful execution of the project will position them for success within the organization. This allows the center to grasp a comprehensive understanding of the client's need, build software solutions that address those needs, and propose future upgrades that will position the client for sustained relevance within the client's organization. 
\paragraph{\textbf{Lesson learned 7.} Iterative project management increases client satisfaction} The experience at the center shows that the iterative project management process works best in meeting the client's expectations. The client is excited and assured of the project's successful completion when they are able to see a minimum viable product during the first scheduled meeting after signing the contracts. This also motivates the student to work harder and successfully implement all the requested features. The rapid development of a minimum viable product or prototype is due to the adoption of iterative project management instead of traditional waterfall practices.
\paragraph{\textbf{Lesson learned 8.} Don't be discouraged to take on a small project} Many of the projects executed by the center started as small projects that cost a few thousand dollars. However, the successful completion of the project and the regular upgrades proposed by the center motivated the client to expand the scope of the project with bigger budgets.
\paragraph{\textbf{Lesson learned 9.} Make sure the University management are involved} The student-driven solution center is a unit under the university and it must conform to the rules and regulations of the university. Hence, the support of the university management is important in sustaining the center. University support may be needed for building facilities, student recruitment, conflict resolution with clients, contract validation, and so on.
\paragraph{\textbf{Lesson learned 10.} As you grow, keep an eye on student empowerment} As a center continues to grow and attracts more clients, it is important to continuously emphasize the need for student empowerment and growth. The successful migration of a project to the production phase with live users means there is less tolerance for errors. Furthermore, there may be fewer opportunities at the top of the cadre for more experienced students who have been with the center for a long time. Hence it is important continuously emphasize the need to always create more opportunities for both new and long-term students at the center to grow and enhance their skills.
\spa{
\section{Limitations and Threats to Validity}
There are a number of threats that may affect the validity of the results discussed in this paper. This section provides an overview of the threats and how we tried to minimize them.
\begin{enumerate}
    \item \textbf{Study Subjects.} The second and third authors were also subjects in the research. This is necessary because majority of the information are from personal observations over the years. Secondly, both authors also contributed significantly to the idea generation, analysis, writing, and review of the paper. Thirdly, since the papers has a lot of historical precedent, the subjects by being authors also accept responsibility for the accuracy of the information presented in the paper. This threat has being minimized with a neutral party (the first author), who lead and moderated the discussions.
    \item \textbf{Student perspective.} The findings reported in this paper has been extracted from  historical data. No study was conducted to validate the findings from the perspective of the students who were interns at the center. However, the archived data at the center shows that 100\% of the student who intern with the center gets a job within 12 months of graduating. This may be an indication of industry-relevant skills being acquired at the center. Furthermore, interns tends to stay for multiple semesters (alternating between part time and full time). This may also be an indication of student satisfaction. A comprehensive study from student perspective is planned for the future.
    \item \textbf{Software Engineering Research.} At the inception, the center was focused on providing value to students and clients. Furthermore, the primary academic unit associated with the center was a teaching-only unit without a PhD program until 2019. Hence, there was limited activities related to software engineering research at the center. For example, there has been no formal study to investigate team dynamics, understand the impact of the center on student skills, or develop novel software engineering processes. This paper represents the first step towards research publications and active research in software engineering. 
    \item \textbf{Geographical location.} Cincinnati is home to three fortune 500 companies and lots of medium-sized companies. The University of Cincinnati is an R1 university and the School of Information Technology currently enrolls more than two thousand students. This provides the center with a large number of potential clients as well as huge pool of student to recruit as interns. Hence, it possible that developing a similar center in another environment (e.g rural areas) may pose different challenges.
\end{enumerate}
}

\section{Conclusion}
This paper introduced the concept of a student-driven software solution center that provides opportunities for students to lead the design, development, execution, and maintenance of software projects for industrial companies. A student-driven solution center that is affiliated with a university can help prepare students to meet the demands of the workforce after graduation and enhance the credibility of the university. The solutions center can also provide an opportunity for faculty to identify practical research problems and align the course contents with emerging industrial demands and technologies. Hence, the establishment of a university-affiliated and student-driven solutions center provides enormous opportunities for students, faculty, and the university.

This paper has presented an experience report on the Information Technology Solution Center, a student-driven solution center affiliated with the School of Information Technology at the University of Cincinnati. The center was established over a decade ago,  has trained over 100 students, and successfully completes about 20 projects annually with industrial partners including Fortune 500 companies and government entities. This paper discussed the needs that initiated the development of the center, the management of the center operations, sample projects that have been successfully completed at the center, and the opportunities and challenges associated with running a student-driven solutions center. These processes have been discussed in detail to provide a clear plan of action for any university willing to establish a similar center.

The establishment of the center was able to satisfy the needs of the students by providing them an opportunity to grow, enhance their technical skills, improve their confidence level, and expose students to projects with real-world impact. The center also satisfies the needs of the faculty by providing them an opportunity to learn new things, identify practical research problems, and earn extra income. The center satisfied the needs of the School of Information Technology by enhancing the school's visibility and credibility and providing external funding to drive the growth of the school. Finally, the center satisfied the needs of the industries by providing a low-cost option to execute important but non-critical projects. It is hoped that this paper will motivate and provide a blueprint for universities to establish a student-driven software solution center.

\newpage
\bibliographystyle{elsarticle-num} 
\bibliography{ref}

\section*{About the authors}
\paragraph{\textbf{Saheed Popoola}} is an assistant professor at the School of Information Technology, University of Cincinnati. He has a PhD in computer science from the University of Alabama. His research interests are in software engineering and computing education. You can contact him at saheed.popoola@uc.edu or visit \url{http://sopopoola.github.io}

\paragraph{\textbf{Vineela Kunapareddi}}
Vineela is an assistant professor-educator at the School of Information Technology, University of Cincinnati. She is also a faculty-fellow at the Information Technology Solutions Center. Vineela holds a Bachelors and Masters degree from the University of Cincinnati and has been part of the ITSC as a student intern, co-op student, graduate assistant, research associate (operations lead), and currently serves as a faculty fellow with the center.

\paragraph{\textbf{Dr. Hazem Said}} is a Professor of Information Technology and the director of the School of Information Technology (SoIT) at the University of Cincinnati (UC). He is a certified Project Management Professional (PMP). Dr. Said founded the UC Information Technology Solutions Center (ITSC) in 2012, where he consults with government, public and private organizations, and leads teams of professionals, as well as graduate, undergraduate, and high school students, to investigate, develop, and support a variety of information technology solutions. In addition, Dr. Said is a co-founder and co-director of the Ohio Cyber Range Institute and the Justice, Law, and Information Technology Institute. Dr. Said is the recipient of over 200 grants and contracts totaling over \$30 million and has authored over 30 articles on topics related to information technology education.

\end{document}